\documentclass[nofootinbib,  preprint]{revtex4-1}
\pdfoutput=1

\usepackage{amsmath}
\usepackage{amssymb}
\usepackage{subfigure}
\usepackage{bm}   
\usepackage{verbatim} 
\usepackage{natbib}

\usepackage{hyperref}
\hypersetup{colorlinks,
 linkcolor = blue,
 anchorcolor = red,
 citecolor = blue,
 filecolor = red,
 urlcolor = red,
 linktocpage}
 
\newcommand{\be}{\begin{equation}}
\newcommand{\ee}{\end{equation}}
\newcommand{\bea}{\begin{eqnarray}}
\newcommand{\eea}{\end{eqnarray}}
\newcommand{\ba}{\begin{align}}
\newcommand{\ea}{\end{align}}

\DeclareSymbolFont{AMSb}{U}{msb}{m}{n}
\DeclareMathSymbol{\N}{\mathbin}{AMSb}{"4E}
\DeclareMathSymbol{\Z}{\mathbin}{AMSb}{"5A}
\DeclareMathSymbol{\R}{\mathbin}{AMSb}{"52}
\DeclareMathSymbol{\Q}{\mathbin}{AMSb}{"51}
\DeclareMathSymbol{\p}{\mathbin}{AMSb}{"50}
\DeclareMathSymbol{\I}{\mathbin}{AMSb}{"49}
\DeclareMathSymbol{\C}{\mathbin}{AMSb}{"43}
\DeclareMathSymbol{\D}{\mathord}{symbols}{"72}
\DeclareMathSymbol{\HH}{\mathbin}{AMSb}{"48}
 
\def\tr{\mathop{\rm Tr}}
\def\d{\partial}
\def\ad{a^\dagger}
\def\D{\Delta}

\def\d{\partial} 
\def\la{\langle}
\def\ra{\rangle}
\def\w{\omega}
\def\half{\textstyle{\frac12}}
\def\ph{\varphi}
\def\phi{\varphi}
\def\tpi{\tilde\pi}
\def\e{\varepsilon}

\begin{document}

\title{Entanglement Entropy and Boundary Conditions in 1+1 Dimensions}

\author{Ben Michel\footnote{\tt michel@physics.ucsb.edu}  and Mark Srednicki\footnote{\tt mark@physics.ucsb.edu}}
\affiliation{Department of Physics, University of California, Santa Barbara, CA 93106 USA}
\date{\today}

\begin{abstract}
Calculations of the entanglement entropy of a spatial region in continuum quantum field theory
require boundary conditions on the fields at the fictitious boundary of the region.
These boundary conditions impact the treatment of the zero modes of the fields
and their contribution to the entanglement entropy. We explore this issue in the simplest example,
the $c=1$ compact-boson conformal field theory in $1{+}1$ dimensions.
We consider three different types of boundary conditions:
spatial Neumann, temporal Neumann, and Dirichlet.
We argue that the first two are well motivated, and show that they lead to the same
result for the Renyi entropies as well as the entanglement entropy,
including a constant term that corresponds to the  Affleck-Ludwig boundary entropy.
The last set of boundary conditions is less well motivated, 
and leads to a different value of the constant term.
The two values are related by a duality transformation on the compact boson.
We also verify some of our results with heat-kernel methods.
\end{abstract}

\maketitle

\setcounter{footnote}{0}

\section{Introduction}
\label{Intro}

Entanglement entropy in quantum field theory is now a well studied subject with
a number of different applications \cite{Calabrese:2004eu,Casini:2009sr,Eisert:2008ur}.
In the case of conformal field theories,
conformal symmetry provides powerful analytical tools for the computation
of entanglement entropy (and, more generally, Renyi entropies) \cite{Casini:2008cr,Casini:2010kt,Casini:2011kv,Klebanov:2011uf}, particularly in 
two spacetime dimensions \cite{Calabrese:2009qy}. 
The abstract nature of the these methodologies, however, can make the underlying
physics obscure. 
In this paper, we do some basic computations with more pedestrian methods,
in an attempt to elucidate some of the underlying issues.

In general, we 
consider a quantum field theory in $d$-dimensional Minkowski space $R^{1,d-1}$
with a unique ground state $|0\ra$.
We choose a finite contiguous spatial region $A$, and construct a density matrix $\rho$ by 
tracing out the fields that live in the complementary region $\overline A$,
\be
\rho={\mathop{\rm Tr}}_{\overline A}|0\ra\la 0|.
\label{rho}
\ee
The Renyi entropy with index $\beta$ is then defined as
\be
S_\beta = {1\over 1-\beta}\log\tr\rho^\beta,
\label{Sb}
\ee
and the entanglement entropy is 
\be
S_{\rm EE}=-\tr\rho\log\rho=\lim_{\beta\to1}S_\beta .
\label{S}
\ee
Even in the simplest case of scalar fields,  this procedure is ambiguous in the continuum,
because in general specifying ``the fields in region $A$'' requires boundary conditions at the
surface of $A$. In the case of gauge fields, much more analysis is needed \cite{Donnelly:2015hxa}, 
since there is in general no gauge-invariant prescription for assigning the gauge potentials to regions \cite{Donnelly:2014gva}. 
For scalar fields in more than two spacetime dimensions, the issue of boundary conditions
has been raised in the context of the influence of the conformal coupling of the scalar fields to the 
background spacetime curvature \cite{Herzog:2016bhv}.
In two spacetime dimensions,
the conformal coupling vanishes and so does not affect the calculation, but the question of appropriate boundary conditions remains \cite{Ohmori:2014eia,Cardy:2016fqc}.
If the conformal field theory in question arises as the continuum
limit of an underlying lattice theory, or we have other information about the
ultraviolet regulator, it can be possible to deduce the appropriate
boundary condition.

Our goal here is to explore the role of boundary conditions at the surface $A$ in
a simple example of a continuum conformal field theory in $1{+}1$ dimensions,
the $c=1$ compact-boson.  
We will find that zero modes play the key role, as they do in gauge theories \cite{Donnelly:2015hxa,Donnelly:2016mlc}.

\section{Review of results from conformal field theory}
\label{rev}

The density matrix $\rho$ of Eq.~(\ref{rho}) is a hermitian operator
with nonnegative eigenvalues that obeys 
\be
\tr\rho=1.
\label{trrho}
\ee
Therefore we can write it in the form
\be
\rho=\exp(- K),
\label{eK}
\ee
where $K$ is a hermitian operator known as the modular hamiltonian \cite{Haag1996}.
In the special case that the quantum field theory is a conformal field theory, 
and the spatial region $A$ is a $(d{-}1)$-dimensional ball of radius $R$,
conformal symmetry can be used to show that, up to a possible boundary term that
will be discussed further below, $K$ is the generator of a conformal
transformation that preserves the boundary of the ball; this leads to \cite{Casini:2008cr,Casini:2011kv}
\be
K=\int_A d^{d-1}x\,f(x)T^{00}(x)+c,
\label{K}
\ee
where $T^{\mu\nu}$ is the conformal traceless stress-energy tensor, obtained by varying the background metric, and including contributions from a coupling $\xi R\ph^2$ to the background curvature $R$, with 
$\xi={d-2\over 4(d-1)}$; the function $f(x)$ is
\be
f(x) ={\pi\over R} (R^2-|x|^2),
\label{fx}
\ee
and $c$ is a constant that is fixed by the requirement $\tr\rho=1$.

The simplest case to consider is a single real compact scalar field $\ph$,
\be
\ph(x) \sim \ph(x)+ V,
\label{phV}
\ee
where $V$ is the circumference of the target-space circle for $\ph$.
The conformal energy density is
\be
T^{00}=\half\pi^2+\half(\d_x\ph)^2 +\xi\nabla^2\ph^2,
\label{T00}
\ee
where $\pi$ is canonically conjugate to $\ph$,
\be
[\ph(x,t),\pi(x',t)]=i\delta(x-x').
\label{comm}
\ee
There has been extensive discussion of whether an extra boundary term is
also needed; see \cite{Herzog:2016bhv} and references therein. This issue arises from the 
coupling to the background curvature, and is absent in $d=2$, where $\xi=0$.
We therefore specialize to this case, 
since the issues we wish to explore arise even in this simplest situation.

Having specialized to $d=2$ and $\xi=0$, we can make a change of the spatial coordinate,
\begin{align}
dy &= {dx\over f(x)}, 
\label{dy} \\
y&={1\over\pi}\tanh^{-1}(x/R).
\label{y} 
\end{align}
The range $x\in(-R,R)$ corresponds to $y\in(-\infty,\infty)$.
We then have
\be
\d_y\ph(y)= f(x)\d_x\ph(x).
\label{dyph}
\ee
If we now define a rescaled conjugate momentum
\be
\tpi(y) = f(x)\pi(x),
\label{tpi}
\ee
then we have
\be
[\ph(y),\tpi(y')]=i\delta(y-y')
\label{commy}
\ee
and
\be
K=\int_{-\infty}^{+\infty}dy\,\left[\half\tpi^2+\half(\d_y\ph)^2\right]+c.
\label{K2}
\ee
Thus the modular hamiltonian corresponds to the hamiltonian of a free-field theory
on an infinite line. 

The entanglement entropy computed from Eqs.~(\ref{S},\ref{eK},\ref{K2}) is infinite.
To regulate it, we give the $y$ coordinate a finite range, $y\in(-L/2,L/2)$, with $L\gg 1$.
This corresponds to a cutoff on the $x$ coordinate at $|x|=R-\e$ with
$\e\ll R$, and 
\be
L={1\over\pi}\log\left({2R\over\e}\right)+O(\e).
\label{L}
\ee
Thus the $\e\to0$ limit corresponds to $L\to \infty$.
It will also be convenient to shift the origin of the $y$ coordinate, so that
$y\in(0,L)$. We now have
\be
K=\int_{0}^{L}dy\,\bigl[\half\tpi^2+\half(\d_y\ph)^2\bigr]+c,
\label{K4}
\ee
which is the hamiltonian of a free-field theory on a finite interval of length $L$. To fully specify $K$, 
we will need to choose boundary conditions at the ends of the interval. 

We begin by writing general mode expansions for $\ph$, $\d_y\ph$, and $\tpi=\d_t\ph$,
\begin{align}
\ph(y,t) &=\ph_0+\ph_1 y +\pi_0 t+\pi_1 yt+\sum_{k=1}^\infty \chi_k(y)\bigl[a_k e^{-i\w_k t}+\ad_k e^{i\w_k t}\bigr],
\label{phyt} \\
\d_y\ph(y,t) &=\ph_1+\pi_1 t+\sum_{k=1}^\infty \chi'_k(y)\bigl[a_k e^{-i\w_n t}+\ad_k e^{i\w_k t}\bigr],
\label{dphyt} \\
\tpi(y,t) &=\pi_0 +\pi_1 y-i\sum_{k=1}^\infty \w_n \chi_k(y)\bigl[a_k e^{-i\w_n t}-\ad_k e^{i\w_k t}\bigr],
\label{piyt} 
\end{align}
where 
\be
[a_k,\ad_{k'}]=\delta_{kk'},
\label{akadk}
\ee
and the mode function $\chi_k(y)$ satisfies $\chi''_k+\w_k^2 \chi_k=0$.
We have assumed that the boundary conditions will render the allowed frequencies discrete and the mode functions real.

Since $\tpi(y)=f(x)\pi(x)$ and $\d_y\ph(y)=f(x)\d_x\ph(x)$, and since $f(x)$ vanishes at the boundary $|x|=R$,
a natural choice of boundary condition is to require both $\tpi(y)$ and $\d_y\ph(y)$ to vanish at the 
regularized boundary points $y=0$ and $y=L$. However, requiring both is incompatible with the 
commutation relation, Eq.~(\ref{comm}). 
So we must make choice. We discuss three possible choices in the next three sections.

\section{Spatial Neumann boundary conditions}
\label{nbc}

We first consider what we will call spatial Neumann boundary conditions,
\be
\d_y\ph(y,t)\big|_{y=0} = 0, \quad \d_y\ph(y,t)\big|_{y=L} = 0.
\label{nbc}
\ee
From Eq.~(\ref{dphyt}), we see that spatial Neumann boundary conditions require 
$\chi'_k(0)=\chi'_k(L)=0$, which
fixes $\chi_k(y)\sim\cos(\w_k y)$ and 
\be
\w_k = \pi k/L.
\label{wk}
\ee
The correctly normalized mode functions are then
\be
\chi_k(y) = {1\over\sqrt{\pi k}}\cos(\pi k y /L).
\label{chin}
\ee
We also have the zero-mode conditions
\be
\ph_1=0,\quad \pi_1=0.
\label{ph1pi1}
\ee
The remaining zero modes are then given by
\be
\ph_0=\int_0^L dy\;\ph(y,t),\quad
\pi_0=\int_0^L dy\;\tpi(y,t),
\label{ph0pi0}
\ee
and from the commutation relation, Eq.~(\ref{comm}), we find
\be
[\ph_0,\pi_0]={i\over L}.
\label{comm0}
\ee
Since $\ph$ is compact, we have from Eq.~(\ref{phV}) that $\ph_0\sim\ph_0+V$, and hence $\pi_0$ is quantized,
\be
\pi_0 = {2\pi\over LV}m,\quad m=0,\pm1,\ldots.
\label{pi0m}
\ee
The modular hamiltonian is now
\be
K = {2\pi^2\over LV^2}m^2 + \sum_{k=1}^\infty \w_k\ad_k a_k+c.
\label{KSN}
\ee
We can now compute
\be
\tr e^{-\beta K}=e^{-\beta c}Z_{\rm osc}Z_{\rm 0,SN}, 
\label{tr1}
\ee
where the oscillator partition function is
\begin{align}
Z_{\rm osc}
&= \prod_{k=1}^\infty \sum_{n=0}^\infty e^{-\beta\w_k n}
\label{zosc1} \\
&= \prod_{k=1}^\infty {1\over 1-e^{-\beta\w_k}}
\label{zosc1a} \\
&=\bigl[e^{\pi\beta/24L}\, \eta(i\beta/2L)\bigr]^{-1},
\label{zosc2}
\end{align}
where $\eta(\tau)$ is the Dedekind eta function.
For large $L$ (more specifically, $L\gg\beta$), we have
\be
\log Z_{\rm osc} = {\pi L\over 6\beta}-{1\over2}\log\!\left({2L\over\beta}\right) + O(\beta/L).
\label{logzosc}
\ee
If we ignore the zero mode completely, and compute the Renyi entropy from
$\tr\rho^\beta=e^{-\beta c}Z_{\rm osc}$ with $c$ adjusted to make $\tr\rho=1$,
we find
\be
S^{\rm osc}_\beta = {(1+\beta)\over 6\beta}\pi L-{1\over2}\log{2L}+{\log\beta\over2(1-\beta)}.
\label{Sosc}
\ee
Recalling that $\pi L=\log(2R/\e)$, we see that the first term is the usual result.
However there is an additional $\log L\sim\log\log R/\e$ term, which is anomalous
and not expected to appear in the final answer. The possibility of such a term was noted in \cite{Donnelly:2015hxa}. This result for the entanglement entropy was first found by \cite{Holzhey:1994we}, where the subleading term was not retained.

This anomalous term is canceled when we include the contribution of the zero modes.
The zero-mode partition function with spatial Neumann boundary conditions is
\begin{align}
Z_{\rm 0,SN}
&= \sum_{m=-\infty}^{+\infty}\!\!\!e^{-2\pi^2\beta m^2/V^2 L}
\label{Z0N1} \\
&=\vartheta(2\pi i\beta/LV^2).
\label{Z0N2}
\end{align}
where $\vartheta(\tau)$ is a Jacobi theta function (with the other argument $z=0$).
For large $L$, we have
\be
\log Z_{\rm 0,SN} = {1\over 2}\log\!\left({2L\over \beta}\right)
+{1\over 2}\log\!\left({V^2\over 4\pi}\right)+O(e^{-LV^2/2\pi\beta}).
\label{logZ0N}
\ee
Adding this to Eq.~(\ref{logzosc}), we have
\be
\log Z_{\rm osc}+\log Z_{\rm 0,SN}=
{\pi\over 6\beta}L+{1\over 2}\log\!\left({V^2\over 4\pi}\right) + O(\beta/L).
\label{logzsum}
\ee
The anomalous $\log L$ has now been canceled. The resulting Renyi entropy is
\be
S_\beta = {\pi(1+\beta)L\over 6\beta}+{1\over 2}\log\!\left({V^2\over 4\pi}\right) + O(\beta/L).
\label{SbN}
\ee
This is the usual result. The constant term, independent of both $L$ and $\beta$ but depending
on $V$, can be understood as a contribution from the Affleck-Ludwig boundary entropy \cite{Affleck:1991tk}
with these boundary conditions \cite{Cardy:2016fqc,Ohmori:2014eia}.

\section{Temporal Neumann boundary conditions}
\label{wdbc}
We next consider what we will call temporal Neumann boundary conditions, 
\be
\tpi(0,t)= 0, \quad \tpi(L,t) = 0.
\label{wdbc1}
\ee
Again, these are motivated by $\tpi(y)=f(x)\pi(x)$, and the vanishing of $f(x)$ at the ends
of the interval in the original $x$ coordinate.
From Eq.~(\ref{dphyt}), we see these boundary conditions require $\chi_k(0)=\chi_k(L)=0$, which
fixes $\chi_k(y)\sim\sin(\w_k y)$ and Eq.~(\ref{wk}).
The correctly normalized mode functions are then
\be
\chi_k(y) = {1\over\sqrt{\pi k}}\sin(\pi k y /L).
\label{shin}
\ee
We also have the zero-mode conditions
\be
\pi_0=0,\quad \pi_1=0.
\label{ph1pi1}
\ee
The remaining zero modes are then $\ph_0$ and $\ph_1$,
with $\ph_0\sim\ph_0+V$; $\ph_1$ has an infinite range.
The modular hamiltonian is now
\be
K = {L\over2}\ph_1^2 + \sum_{k=1}^\infty \w_k\ad_k a_k+c.
\label{KTN}
\ee
The oscillator contribution is therefore the same as it is for spatial
Neumann boundary conditions,
Eq.~(\ref{zosc2}).
The zero-mode contribution is now
\be
Z_{\rm 0,TN}
= L\int_0^V d\ph_0\int_{-\infty}^{+\infty}d\ph_1\,e^{-\beta L\ph_1^2/2}
\label{Z0WD1}
\ee
The prefactor of $L$ in Eq.~(\ref{Z0WD1}) arises from the measure for a trace of a functional of $\ph(x)$,
\be
\prod_x d\ph(x) = L\,d\ph_0\,d\ph_1\prod_{n=1}^\infty c_n,
\label{dph}
\ee
where $c_n$ is the coefficient of $\chi_n(x)$ in the mode expansion of $\ph(x)$.
The factor of $L$ comes from the jacobian for this change of integration variables;
its necessity can be seen from dimensional analysis.
Evaluating the integral in Eq.~(\ref{Z0WD1}), we have
\be
Z_{\rm 0,TN}
=V\biggl({2\pi L\over\beta }\biggr)^{\!\!1/2}.
\label{Z0WD2}
\ee
This yields
\be
\log Z_{\rm 0,TN} = {1\over 2}\log\!\left({2L\over \beta}\right)
+{1\over 2}\log\!\left({V^2\over 4\pi}\right)+O(e^{-LV^2/2\pi\beta}).
\label{logZ0WD}
\ee
For large $L$, this is the same as $\log Z_{\rm 0,SN}$,
Eq.~(\ref{logZ0N}), 
up to exponentially small corrections, and therefore the result for the Renyi
entropy is also the same, Eq.~(\ref{SbN}).

\section{Dirichlet boundary conditions}
\label{sdbc}
Although not motivated by the vanishing of $f(x)$ at the endpoints, in this section
we consider Dirichlet boundary conditions
\be
\ph(0,t)= 0, \quad \ph(L,t) = 0 \mathrel{\rm mod} V.
\label{sdbc}
\ee
From Eq.~(\ref{dphyt}), we see these conditions require $\chi_k(0)=\chi_k(L)=0$, which
results in Eqs.~(\ref{wk}) and (\ref{shin}), the same as for temporal Neumann boundary conditions.
The zero-mode conditions are now
\be
\ph_0=0,\quad \pi_0=0,\quad \pi_1=0.
\label{ph0ph1pi1}
\ee
The remaining zero mode is then
\be
\ph_1 = {wV\over L}, \quad w=0,\pm1,\ldots
\label{ph1wv}
\ee
The modular hamiltonian is now
\be
K = {V^2\over2L}w^2 + \sum_{k=1}^\infty \w_k\ad_k a_k+c.
\label{KD}
\ee
The oscillator contribution is therefore the same as it is for spatial or temporal 
Neumann boundary conditions,
Eq.~(\ref{zosc2}). The zero-mode contribution is now
\begin{align}
Z_{\rm 0,D}
&= \sum_{w=-\infty}^{+\infty}\!\!\!e^{-\beta V^2 m^2/2 L}
\label{Z0SD1} \\
&=\vartheta(i\beta V^2/2\pi L).
\label{Z0SD2}
\end{align}
For large $L$, we have
\be
\log Z_{\rm 0,D} = {1\over 2}\log\!\left({2L\over \beta}\right)
-{1\over 2}\log\!\left({V^2\over \pi}\right)+O(e^{-2\pi L/\beta V^2}).
\label{logZ0SD}
\ee
Combining Eqs.~(\ref{logzosc}) and (\ref{logZ0SD}) for large $L$, we have
\be
\log Z_{\rm osc}+\log Z_{\rm 0,D}=
{\pi L\over 6\beta}-{1\over 2}\log\!\left({V^2\over \pi}\right) + O(\beta/L).
\label{logzsum2}
\ee
The anomalous $\log L$ has now again been canceled. The resulting Renyi entropy is
\be
S_\beta = {(1+\beta)\over 6\beta}\pi L-{1\over 2}\log\!\left({V^2\over \pi}\right) + O(\beta/L).
\label{SbSD}
\ee
The constant term is different than it is in the case of spatial or temporal Neumann boundary conditions,
and again can be understood as a contribution from the Affleck-Ludwig boundary entropy \cite{Affleck:1991tk}.

\section{Duality}
\label{duality}
Comparing Eq.~(\ref{SbN}) for the Renyi entropy with Neumann boundary conditions (spatial or temporal)
with Eq.~(\ref{SbSD}) for the Renyi entropy with Dirichlet boundary conditions, we see
that they are related by
\be
V \leftrightarrow {2\pi\over V}.
\label{V2}
\ee
For the case of spatial Neumann boundary conditions, this follows from the same relation 
for the zero-mode partition functions, Eqs.~(\ref{Z0N2}) and (\ref{Z0SD2}).
This is related to the $T$-duality transformation for the compact boson on a spatial circle
with circumference $2L$ with periodic boundary conditions. In this case, the mode expansion is
\be
\ph(y,t)=\ph_0+\ph_1 y + \pi_0 t 
+{1\over\sqrt{2\pi}}\sum_{k=1}^\infty {1\over k}
\bigl[a_k e^{-i\w_k(y-t)}+\tilde a_k e^{i\w_k(y+t)}+\hbox{h.c.}\bigr],
\label{phyt2}
\ee
with
\be
\ph_1= {V\over 2L}w, \quad \pi_0 = {\pi\over LV}m,
\label{ph1pi0}
\ee
where again $w$ and $m$ are integers representing winding and momentum modes, and
now there are two types of oscillators (left moving and right moving modes) with $\w_k=\pi k/L$;
matching Eq.~(\ref{wk}) is the reason for having the circle be twice as long as the interval.
The hamiltonian is
\be
H = {\pi^2\over LV^2}m^2 + {V^2\over 4L}w^2 
+\sum_{k=1}^\infty \w_k\bigl(\ad_k a_k+\tilde a_k^\dagger \tilde a_k\bigr)+c.
\label{H}
\ee
With a judicious choice of $c$, and introducing the modular parameter 
\be
\tau = i\beta/2L,
\label{tau}
\ee
the complete partition function is
\be
Z_{\rm circle}(\tau,V) 
= {\vartheta(2\pi \tau/V^2)\vartheta(\tau V^2/2\pi )\over\eta(\tau)^2} 
\label{zc1} 
\ee
which is manifestly invariant under Eq.~(\ref{V2}). We also have the relation \cite{Eggert:1992ur}
\be
Z_{\rm circle}(\tau,V) 
= Z_{\rm SN}(\tau,\sqrt{2}V)Z_{\rm D}(\tau,V/\sqrt{2})
\label{zc2}
\ee
for particular choices of $c$ in Eqs.~(\ref{KSN}) and (\ref{KD}).
Since $Z_{\rm circle}(\tau,V)$ 
can be written as a euclidean path integral over the field on a 2-torus with
cycle lengths $\beta$ and $2L$, we also have invariance under $\beta\leftrightarrow 2L$
or equivalently $\tau\leftrightarrow -1/\tau$, which follows from
\begin{align}
\eta(-1/\tau) &= (-i\tau)^{1/2}\eta(\tau),
\label{etadual} \\
\vartheta(-1/\tau) & =(-i\tau)^{1/2}\vartheta(\tau).
\label{thetadual}
\end{align}
The boundary conditions on the interval of length $L$ break this symmetry, but as a remnant
of it we have the relations
\begin{align}
Z_{\rm SN}(-1/\tau,V) &= \sqrt{V^2/4\pi} \, Z_{\rm D}(\tau,V/2),
\label{zsndual} \\
Z_{\rm D}(-1/\tau,V) &= \sqrt{\pi/V^2} \, Z_{\rm SN}(\tau,2V).
\label{zsndualz} 
\end{align}

\section{Comparison with heat kernel methods}
\label{hksection}

In this section we evaluate the entanglement entropy using heat-kernel methods to facilitate comparison with the computation of Casini and Huerta \cite{Casini:2010kt}. For a review of these techniques, see \cite{Vassilevich:2003xt}.

The key formula is
\be
\log Z_\text{osc} = \frac{1}{2}\lim_{s\to 1} \int_0^{\infty} \frac{dt}{t^s} K(t),
\label{hk}
\ee
where $K(t) = \tr' e^{-t\D}$
and $\D$ is minus the Laplacian
on the target manifold (in our case, an interval $I$ of length $L$ with one of our sets of boundary conditons) 
times a circle $S^1$ of circumference $\beta$, and the prime on the trace indicates that we omit the zero modes on $I$. Thus we have $\D=-\d_\tau^2-\d_y^2$, where $\tau$ is 
periodic with period $\beta$. Since this is a product space, the trace factorizes,
\be
K(t) = K_{S^1}(t)K_I(t)
\label{kt}
\ee
where $K_{S^1}(t)$ is the regulated form (defined shortly) of the unregulated heat kernel $\tilde K_{S^1}(t)$
on a circle of circumference $\beta$,
\begin{align}
\tilde K_{S^1}(t) &= \sum_{n=-\infty}^{\infty} e^{t(2\pi i n/\beta)^2}
\label{ks1} \\
&= \vartheta(4\pi i t/\beta^2)
\label{ks12} \\
&= \sqrt{\beta^2/4\pi t}\,\vartheta(i\beta^2/4\pi t)
\label{ks13} \\
&= {\beta\over\sqrt{4\pi t}}\sum_{n=-\infty}^{\infty}e^{-n^2\beta^2/4t}.
\label{ks14}
\end{align}
The regulated heat kernel is obtained by dropping the $n=0$ term, or equivalently, subtracting the
$\beta\to\infty$ limit:
\be
K_{S^1}(t)={2\beta\over\sqrt{4\pi t}}\sum_{n=1}^{\infty}e^{-n^2\beta^2/4t}.
\label{ks15}
\ee
We can check that this regulated heat kernel on the circle gives the correct answer for the
partition function by noting that, for a partition function
with the general form of Eq.~(\ref{zosc1a}),
\be
Z = \prod_{k=1}^\infty {1\over 1-e^{-\beta\w_k}},
\label{Z}
\ee
we have
\be
-\log(1-e^{-\beta\w_k})=\frac{1}{2}\lim_{s\to 1} \int_0^{\infty} \frac{dt}{t^s} K_{S^1}(t)e^{-t\w_k^2}.
\label{log}
\ee
Summing over $k$ then gives $\log Z$ on the left, and we identify 
\be
K_I(t)=\sum_{k=1}^\infty e^{-t\w_k^2}
\label{KIt}
\ee
on the right. 

For all our choices of boundary conditions, we have $\w_k=\pi k/L$, and hence
\begin{align}
K_I(t) &= \sum_{k>0} e^{-\pi^2 k^2 t/L^2}
\label{KIt1} \\
&= {1\over2}\left[\vartheta(i\pi t/L^2)-1\right]
\label{KIt2} \\
&= {1\over2}\left[\sqrt{L^2/\pi t}\,\vartheta(iL^2/\pi t)-1\right]
\label{KIt3} \\
&= {1\over2}\left[\sqrt{L^2/\pi t}\left(2\sum\nolimits_{m>0} e^{-m^2 L^2/t}+1\right)\right]-1
\label{KIt4} \\
&= \frac{L}{\sqrt{4\pi t}}\left[1+ 2\sum_{m>0} e^{-m^2 L^2/t} - \frac{\sqrt{\pi t}}{L}\right].
\label{KIt5}
\end{align}
Plugging back into Eq.~(\ref{hk}), we find
\begin{align}
\log Z_{\rm osc} &= {\beta L\over4\pi}\lim_{s\to 1} \int_0^{\infty} \frac{dt}{t^{s+1}} 
\sum_{n>0} e^{-n^2\beta^2/4t} \left[1+ 2\sum_{m>0} e^{-m^2 L^2/t} - \frac{\sqrt{\pi t}}{L}\right]\nonumber\\
&= I_1 + I_2 + I_3.
\label{I123}
\end{align}
In the analysis of \cite{Casini:2010kt}, only $I_1$ is kept; the second and third terms 
are subleading in $L$, and are dropped. However, $I_2$ and $I_3$ must also be included
in order to reproduce the canonical oscillator partition function, Eq.~\eqref{zosc2}, and supplemented by the appropriate zero-mode contribution to obtain the entropy.
We compute the integrals in Appendix \ref{appdx:integrals}, with the result
\begin{align}
I_1 &= \frac{\pi L}{6\beta},\nonumber\\
I_2 &=-\frac{\pi L}{6\beta}-\frac{\pi \beta}{24L} -\log\eta(i\beta/2L)
+\frac{1}{4}\lim_{s\to 1}\left(\frac{1}{s-1}+\gamma_E-2\log \beta\right),\nonumber\\
I_3 &= -\frac{1}{4}\lim_{s\to 1}\left(\frac{1}{s-1}+\gamma_E-2\log\beta\right)
\end{align}
where $\gamma_E$ is the Euler-Mascheroni constant. Thus the heat kernel gives
\be
\log Z_{\rm osc} = I_1+I_2+I_3 = -\frac{\pi \beta}{24L} -\log\eta(i\beta/2L)
\label{zosc3}
\ee
in exact agreement with Eq.~\eqref{zosc2}.

\section{Discussion}

We have undertaken a detailed analysis of the computation of the entanglement entropy
(and the Renyi entropies) of an interval for the compact-boson $c=1$ conformal field theory
in $1{+}1$ dimensions, paying particular attention to the role of boundary conditions at the endpoints
and the contributions of zero modes, and using operator methods rather than euclidean path
integral manipulations. The possibility of temporal Neumann boundary conditions, and the
distinction of them from Dirichlet boundary conditions, does not seem to have been previously
considered. Our results emphasize the necessity of paying careful attention to these issues.

\acknowledgments

We are happy to thank Vlad Rosenhaus, Dan Harlow, and especially 
Will Donnelly and Aron Wall for extensive discussions.
This work was supported in part by NSF Grant PHY13-16748.

\appendix

\section{Integrals from \S\ref{hksection}}
\label{appdx:integrals}
In this appendix we evaluate the integrals in \eqref{I123}. 
\begin{align}
I_1 &=\frac{\beta L}{4\pi } \lim_{s\to 1} \sum_{n>0} \int_0^{\infty} \frac{dt}{t^{s+1}} e^{-n^2\beta^2/4t} 
\nonumber \\
&= \frac{\beta L}{4\pi} \lim_{s\to 1} 4^{s} \beta^{-2s}\Gamma(s)\zeta(2s) 
\nonumber \\
&= \frac{\pi L}{6\beta}.
\label{I1}
\end{align}
Note that we could have set $s=1$ at the start for $I_1$, but the same is not true for $I_2$ and $I_3$.
We have
\begin{align}
I_3 &= -{\beta\over4\sqrt{\pi}} \lim_{s\to 1}\int_0^{\infty} \frac{dt}{t^{s+1/2}}
\sum_{n>0} e^{-n^2\beta^2/4t}
\nonumber\\
&= -\frac{\beta}{4\sqrt{\pi}} \lim_{s\to 1}(\beta/2)^{1-2s}\Gamma(s-1/2)\zeta(2s-1) 
\nonumber\\
&= -\frac{1}{4}\lim_{s\to 1}\left(\frac{1}{s-1}+\gamma_E-2\log\beta\right).
\end{align}
Finally, we evaluate $I_2$:
\begin{align}
I_2 &= {\beta L\over 2\pi}\lim_{s\to 1}\int_0^{\infty} \frac{dt}{t^{s+1}}
\sum_{m,n>0} e^{-n^2\beta^2/4t-m^2 L^2/t}
\nonumber\\
&= {\beta L\over 2\pi}\lim_{s\to 1}\Gamma(s)L^{-2s}
\sum_{m,n>0} \frac{1}{[(n\beta/2L)^2+m^2]^s}.
\label{I2}
\end{align}
The double sum can be expressed in terms of an Eisenstein series
\be
E(\tau,s) = \sum_{(m,n)\ne(0,0)}{(\mathop{\rm Im}\tau)^s\over|n\tau+m|^{2s}}
\label{E}
\ee
with $\tau=i\beta/2L$. We have
\begin{align}
\sum_{m,n>0} \frac{1}{[(n\beta/2L)^2+m^2]^s} 
&={1\over 4}(\beta/2L)^{-s}E(i\beta/2L,s)
-{1\over2}(\beta/2L)^{-2s}\sum_{n>0}{1\over n^{2s}}
-{1\over2}\sum_{m>0}{1\over m^{2s}}
\nonumber \\
&={1\over 4}(\beta/2L)^{-s}E(i\beta/2L,s)
-{1\over2}(\beta/2L)^{-2s}\zeta(2s)
-{1\over2}\zeta(2s).
\label{sum2}
\end{align}
Here the sum over the positive quadrant of $\Z^2$ was rewritten as the sum over $(m,n)\neq (0,0)$, minus the $(m,0)$ and $(0,n)$ lines, all divided by 4. We now have
\be
I_2 = {\beta L\over 8\pi}\lim_{s\to 1}\Gamma(s)\left[
(\beta L/2)^{-s} E(i\beta/2L,s)-2(\beta/2)^{-2s}\zeta(2s)-2L^{-2s}\zeta(2s)\right].
\label{I21}
\ee
The limits of the last two terms are simple to evaluate, and yield $-\pi L/6\beta$ and
$-\pi\beta/24 L$, respectively. To evaluate the first term, we need the 
Kronecker limit formula for Eisenstein series near $s=1$,
\be
E(\tau,s) = \frac{\pi}{s-1}+2\pi\left[\gamma_E-\log 2
-\log\left(\sqrt{\mathop{\rm Im}\tau}\,|\eta(\tau)|^2\right)\right]+O(s-1).
\label{Kron}
\ee
With this we find
\be
I_2 = -\log\eta(i\beta/2L) + \frac{1}{4}\lim_{s\to 1}\left(\frac{1}{s-1}+\gamma_E-2\log\beta\right)
-{\pi L\over 6\beta}-{\pi\beta\over 24L}.
\label{I22}
\ee

\bibliographystyle{toine}
\bibliography{eebc_refs.bib}

\end{document}